\documentclass[aps,prl,showpacs,amsmath,amsfonts,amssymb]{revtex4} 
\usepackage[dvips]{graphicx}
\usepackage{bbm}
\usepackage{amsthm}
\usepackage{multirow}

\newcommand{\beq}{\begin{equation}} 
\newcommand{\eeq}{\end{equation}}
\newcommand{\bea}{\begin{eqnarray}} 
\newcommand{\eea}{\end{eqnarray}}

\newtheorem*{conj}{Conjecture}

\input epsf 
 
\begin{document} 
 
\title{On a microcanonical relation between continuous and discrete spin models} 

\author{Lapo Casetti} 
\email{lapo.casetti@unifi.it} 
\affiliation{Dipartimento di Fisica e Astronomia and CSDC, Universit\`a di 
Firenze, and Istituto Nazionale di Fisica Nucleare (INFN), Sezione di
Firenze, via G.~Sansone 1, I-50019 Sesto Fiorentino (FI), Italy}  
\author{Cesare Nardini} 
\email{cesare.nardini@gmail.com} 
\affiliation{Dipartimento di Fisica e Astronomia and CSDC, Universit\`a di 
Firenze, and Istituto Nazionale di Fisica Nucleare (INFN), Sezione di
Firenze, via G.~Sansone 1, I-50019 Sesto Fiorentino (FI), Italy}  
\author{Rachele Nerattini} 
\email{r.nerattini@fastwebnet.it} 
\affiliation{Dipartimento di Fisica e Astronomia and CSDC, Universit\`a di 
Firenze, and Istituto Nazionale di Fisica Nucleare (INFN), Sezione di
Firenze, via G.~Sansone 1, I-50019 Sesto Fiorentino (FI), Italy}  

\date{\today} 
 
\begin{abstract} A relation between a class of stationary points of the energy landscape of continuous spin models on a lattice and the configurations of a Ising model defined on the same lattice suggests an approximate expression for the microcanonical density of states. Based on this approximation we conjecture that if a $O(n)$ model with ferromagnetic interactions on a lattice has a phase transition, its critical energy density is equal to that of the $n = 1$ case, i.e., a system of Ising spins with the same interactions. The conjecture holds true in the case of long-range interactions. For nearest-neighbor interactions, numerical results are consistent with the conjecture for $n=2$ and $n=3$ in three dimensions. For  $n=2$ in two dimensions ($XY$ model) the conjecture yields a prediction for the critical energy of the Bere\v{z}inskij-Kosterlitz-Thouless transition, which would be equal to that of the two-dimensional Ising model. We discuss available numerical data in this respect.
\end{abstract} 
 
 
\maketitle 
The stationary points of a function of many variables are the points of vanishing gradient and play a relevant role in quite a few theoretical methods in physics. When the function is the energy of a many-body system these methods are referred to as ``energy landscape methods'' \cite{Wales}. Examples of applications include clusters \cite{Wales}, disordered systems and glasses \cite{glasses}, biomolecules and protein folding \cite{proteins}. Energy landscape methods allow to estimate dynamic as well as static properties. As far as equilibrium is concerned, the classic application is Stillinger and Weber's thermodynamic formalism \cite{Stillinger}, where minima are the stationary points to be considered. Later, all the stationary points of the energy, including saddles of any index\footnote{The index of a stationary point $p$ of a function $f$ is the number of unstable directions, i.e., the number of negative eigenvalues of the Hessian of $f$ at $p$.} have been taken into account, for instance to characterize glassy behavior \cite{saddles}. It was further realized that stationary points of the Hamiltonian are connected with topology changes of the phase space accessible to the system, leading to the conjecture that some of them are at the origin of thermodynamic phase transitions \cite{PhysRep,JSP}; quite some research activity followed, reviewed in \cite{KastnerRMP,PettiniBook}. 
Although in equilibrium statistical mechanics phase transitions in systems with non-fluctuating particle numbers have been mainly studied within the canonical ensemble, the relation between stationary points of the Hamiltonian and equilibrium statistical properties is more transparent in a microcanonical setting. 
This can be understood in an intuitive way by observing that the entropy is defined as\footnote{Throghout the paper we set Boltzmann's constant $k_B$ to unity and we consider thermodynamic quantities per degree of freedom, whence the $\frac{1}{N}$.} $s(\varepsilon) = \frac{1}{N}\log \omega(\varepsilon)$, where $\varepsilon = E/N$ is the energy density and $\omega$ is the density of states. For a system with $N$ degrees of freedom described by continuous variables the latter can be written as 
\beq
\omega(\varepsilon) = \int_{\Gamma} \delta({\cal H} - N\varepsilon) \, d\Gamma = \int_{\Gamma \cap \Sigma_{\varepsilon}} \frac{d\Sigma}{\left|\nabla {\cal H}\right|}\, ,
\label{coarea}
\eeq
where $\Gamma$ is phase space and $d\Gamma$ its volume measure, $\Sigma_{\varepsilon}$ is the hypersurface of constant energy $E = N\varepsilon$, and $d\Sigma$ stands for the $N-1$-dimensional Hausdorff measure. The rightmost integral stems from a coarea formula \cite{Federer}. At a stationary point, $\nabla{\cal H}=0$ and the integrand diverges, so that its contribution to $\omega$ is clearly important. Indeed,  
the density of states is nonanalytic at stationary values of the energy for any finite $N$, and so is the entropy, at variance with the canonical free energy which may develop nonanalyticities only in the thermodynamic limit $N\to\infty$ \cite{Griffiths}. Microcanonical nonanalyticities at finite $N$ are in one-to-one correspondence with stationary configurations; however, the ``strength'' of such nonanalyticities generically decreases linearly with $N$, i.e., the first $k$ derivatives of the entropy are continuous, where $k$ is ${\cal O}(N)$ (KSS theorem  \cite{KSS,CasettiKastnerNerattini}). The usual thermodynamic functions are given by low-order derivatives of the entropy, so that these nonanalyticities can be observed only for very small $N$ from noisy data. In the thermodynamic limit most of these nonanalyticities disappear. Only those singularities (if any) that survive (or appear) in that limit are typically associated to thermodynamic phase transitions and coincide with the canonical nonanalyticities if equivalence of statistical ensembles holds. This may suggest that finite-$N$ nonanalyticities are totally unrelated to thermodynamic phase transitions. However, this is not true, since Franzosi and Pettini showed that stationary points are a necessary condition for phase transitions to occur, at least for systems with short-range interactions \cite{FranzosiPettini}. Moreover, a possible scenario allowing some finite-$N$ singulartities to survive in the thermodynamic limit has been depicted in \cite{KSS}. Therefore, the relation between the energy landscape, i.e., the stationary points of the Hamiltonian, and thermodynamic phase transitions is still an open problem.

The purpose of the present paper is to apply an energy landscape analysis to $O(n)$ spin models. Loosely speaking, we take seriously the idea that the stationary points contribution to the density of states is the most important one, and by considering a particular class of stationary points we construct an approximate form of the density of states which suggests a relation between the phase transitions occurring in these models. 
Let us consider a classical isotropic spin model defined on a lattice (or more generally on a graph) with Hamiltonian
\beq
{\cal H}^{(n)} = - \sum_{i,j=1}^N J_{ij} S_i \cdot S_j= - \sum_{i,j=1}^N J_{ij} \sum_{\alpha = 1}^n S^a_i S^a_j~,
\label{H}
\eeq
where $i$ and $j$ run over the $N$ lattice sites and the classical spin vectors $S_i = (S_i^1,\ldots,S_i^n)$ have unitary norm, i.e., $\sum_{a = 1}^n \left( S^a_i \right) ^2 = 1$ $\forall i = 1,\ldots,N$. The real matrix $J_{ij}$ dictates the interactions; in case they are long-ranged a normalization is understood such as to obtain an extensive energy, using e.g.\ the Kac prescription \cite{longrangereport}. The Hamiltonian (\ref{H}) is globally invariant under the $O(n)$ group. In the special cases $n=1$, $n=2$, and $n=3$, one obtains the Ising, $XY$, and Heisenberg models, respectively.
The case $n=1$ is even more special because $O(1) \equiv \mathbb{Z}_2$ is a discrete symmetry group. In this special case the Hamiltonian (\ref{H}) becomes the Ising Hamiltonian 
\beq
{\cal H}^{(1)} = - \sum_{i,j=1}^N J_{ij} \sigma_i \sigma_j~,
\label{H_1}
\eeq
where $\sigma_i = \pm 1$ $\forall i$. In all the other cases $n \geq 2$ the $O(n)$ group is continuous; each spin vector $S_i$ lives on an $n-1$ unit sphere $\mathbb{S}_1^{n-1}$. Let us now consider the stationary configurations of ${\cal H}^{(n)}$ for $n\geq 2$, i.e., the solutions $\overline{S} = (\overline{S}_1,\ldots, \overline{S}_N)$ of the $N$ vector equations $\nabla {\cal H}^{(n)} = 0$. The latter can be written as $nN$ scalar equations,
\beq
- \sum_{j=1}^N J_{kj} S^a_j + \lambda_k S^a_k = 0~,  \quad a=1,\ldots,n,~ k = 1,\ldots,N\, ,
\label{eqstationary} 
\eeq
where the $\lambda$'s are $N$ Lagrange multipliers,
plus the $N$ nonlinear constraints $\sum_{a = 1}^n \left( S^a_i \right) ^2 = 1$, which prevent the above equations from being easily solved. However, a particular class of solutions can be found by assuming that all the spins are parallel or antiparallel: $S^1_i = \cdots = S^{n-1}_i = 0$ $\forall i$. In this case, the $N(n-1)$ equations (\ref{eqstationary}) with $a = 1,\ldots,n-1$, corresponding to the first $n-1$ components of the spins, are trivially satisfied. As to the $n$-th component, the constraints $(S^n_i)^2 = 1$ imply $S^n_i = \sigma_i$ $\forall i$, so that the remaining $N$ equations read as
\bea
- \sum_{j=1}^N J_{kj} \sigma_j + \lambda_k \sigma_k = 0~,  \qquad k = 1,\ldots,N\,.
\label{eqstationary_ising} 
\eea
The above equations are satisfied by any of the $2^N$ possible choices of the $\sigma$'s provided one puts $\lambda_k = \left(\sum_{j=1}^N J_{kj} \sigma_j\right)/\sigma_k$, $k = 1,\ldots,N$. The Hamiltonian (\ref{H}) becomes the Ising Hamiltonian (\ref{H_1}) when the spins belong to this class of stationary configurations. Therefore we have a one-to-one correspondence between a class of stationary configurations of the Hamiltonian (\ref{H}) of a $O(n)$ spin model and all the configurations of the Ising model (\ref{H_1}), i.e., the Ising model defined on the same graph with the same interaction matrix $J_{ij}$; the corresponding stationary values are just the energy levels of this Ising Hamiltonian. We shall refer to the class of stationary configurations $\overline{S}_i = (0,\ldots,0,\sigma_i) ~  \forall i = 1,\ldots,N$ as ``Ising stationary configurations''. 
There will be also other stationary configurations; nonetheless, the $2^N$ Ising ones are a non-negligible fraction of the whole, especially at large $N$ because the number of stationary points of a generic function of $N$ variables is expected to be ${\cal O}(e^N)$ \cite{Schilling}. 

The above results hold for $O(n)$ and Ising models defined on any graph. From now on we shall restrict to regular $d$-dimensional hypercubic lattices and to ferromagnetic interactions $J_{ij} > 0$. In this case, in the thermodynamic limit $N\to \infty$ the energy density levels of the Ising Hamiltonian (\ref{H_1}), ${\cal H}^{(1)}(\sigma_1,\ldots,\sigma_N)/N ~ \forall \sigma_i = \pm 1$, become dense and cover the whole energy density range of all the $O(n)$ models. This suggests that Ising stationary configurations are the most important ones, so that we may approximate the density of states $\omega^{(n)}(\varepsilon)$ of an $O(n)$ model in terms of these configurations. To this end, let us first rewrite Eq.\ (\ref{coarea}) as 
\beq
\omega^{(n)}(\varepsilon) = \sum_p \int_{U_p\cap\Sigma_{\varepsilon}} \frac{d\Sigma}{\left|\nabla {\cal H}^{(n)}\right|}
\label{coarea_part}
\eeq
where $p$ runs over the $2^N$ Ising stationary configurations and $U_p$ is a neighborhood of the $p$-th Ising configuration such that $\left\{U_p\right\}_{p=1}^{2^N}$ is a proper partition of the configuration space $\Gamma = \left(\mathbb{S}^{n-1}\right)^N$, that coincides with phase space for spin models (\ref{H}). Since Ising configurations are isolated points in the configuration space of a $O(n)$ model, such a partition always exists. 
Let us now introduce two assumptions allowing to write Eq.\ (\ref{coarea_part}) in a more transparent, albeit approximate, way. $(i)$ At a given value of $\varepsilon$, the largest contribution to $\omega^{(n)}(\varepsilon)$ is likely to come from those $U_p$ such that ${\cal H}^{(n)}(p) = N\varepsilon$, because if ${\cal H}^{(n)}(q) \not= N\varepsilon$ then $\left|\nabla {\cal H}^{(n)}(x)\right| \not= 0$ $\forall x \in U_q \cap \Sigma_\varepsilon$, unless a zero in $\left|\nabla {\cal H}^{(n)}(x)\right|$ comes from a stationary configuration which does not belong to the Ising class. According to our previous considerations, we assume that non-Ising stationary configurations can be neglected. We shall therefore consider only stationary configurations at energy density $\varepsilon$ in the sum (\ref{coarea_part}). $(ii)$ We shall assume that the integrals in Eq.\ (\ref{coarea_part}) depend only on $\varepsilon$, i.e., the neighborhoods $U$ can be deformed such as
\beq
\int_{U_{p}\cap\Sigma_{\varepsilon}} \frac{d\Sigma}{\left|\nabla {\cal H}^{(n)}\right|} = \int_{U_{q}\cap\Sigma_{\varepsilon}} \frac{d\Sigma}{\left|\nabla {\cal H}^{(n)}\right|} = g^{(n)}(\varepsilon)  
\label{g}
\eeq
for any $p,q$ such that ${\cal H}^{(n)}(p) = {\cal H}^{(n)}(q) = N\varepsilon$. Hence, using assumptions $(i)$ and $(ii)$, Eq.\ (\ref{coarea_part}) becomes
\beq
\omega^{(n)}(\varepsilon) \simeq g^{(n)}(\varepsilon)\sum_{p} \delta\left[{\cal H}^{(n)}(p) - N\varepsilon \right] \, .
\label{omega_appr_int}
\eeq 
The sum on the r.h.s. of Eq.\ (\ref{omega_appr_int}) is over Ising configurations, so that it equals the density of states of the corresponding Ising model, that we shall denote by  $\omega^{(1)}(\varepsilon)$. We can thus write 
\beq
\omega^{(n)}(\varepsilon) \simeq  \omega^{(1)}(\varepsilon) \,g^{(n)}(\varepsilon)\, .
\label{omega_appr}
\eeq 
Were Eq.\ (\ref{omega_appr}) exact, it would imply that if $\omega^{(1)}(\varepsilon)$ is nonanalytic at $\varepsilon = \varepsilon_c$, then also $\omega^{(n)}(\varepsilon)$ is nonanalytic at $\varepsilon = \varepsilon_c$ for any $n$, unless the function $g^{(n)}(\varepsilon)$ precisely cancels this nonanalyticity, which seems a rather special case. We do not expect Eq.\ (\ref{omega_appr}) to be exact, even in the thermodynamic limit $N\to\infty$, unless, again, $g(\varepsilon)$ has some very special features: with a generic $g(\varepsilon)$ a density of states of the form (\ref{omega_appr}) would not reproduce the known critical exponents of the $O(n)$ universality classes \cite{Pelissetto_report}. 
However, it can be shown that with a generic $g(\varepsilon)$ Eq.\ (\ref{omega_appr}) correctly implies a negative value for the specific heat critical exponent of $O(n)$ spin models (i.e., the specific heat of continuous models does not diverge at criticality, but rather has a cusp-like behavior). This is a common feature of $O(n)$ models \cite{Pelissetto_report} and reinforces the belief that the approximation (\ref{omega_appr}), although rather crude, may properly capture the main features of the nonanalyticities of the density of states when $N\to\infty$, as the location of such nonanalyticies. 
Therefore we put forward the following
\begin{conj}
If a $O(n)$ spin model defined on a $d$-dimensional hypercubic lattice with Hamiltonian (\ref{H}) and ferromagnetic interaction matrix $J_{ij} >0$ has a phase transition, its critical energy density $\varepsilon_c = E_c/N$ is equal to that of the $n = 1$ case, i.e., a system of Ising spins with the same interactions. 
\end{conj}
The above conjecture concerns the critical value of the control parameter of the microcanonical ensemble, the energy density, and says nothing about critical temperatures, which may well be different---and typically are--- 
at different $n$.
\begin{table}
\caption{Comparison of critical energy densities $\varepsilon_c$ and critical temperatures $T_c$ for ferromagnetic models with long-range (LR) interactions (first row) and nearest-neighbor interactions on a $d$-dimensional hypercubic lattice (all the other rows).}
\begin{ruledtabular}
\begin{tabular}{ccccc}
\multicolumn{2}{c}{model}
& $\varepsilon_c$ & $T_c$ & derivation method\\
 \hline
\multirow{2}{*}{LR\footnote{Long-range interactions $J_{ij} \propto |i-j|^{-\alpha}$ with ${0 \leq \alpha < d}$.}} & Ising  & 0 & 1 & exact solution \\
                            & $O(n)$ & 0 & $1/n$ &  exact solution \cite{Campa} \\ 
 \hline
\multirow{2}{*}{$d=1$} & Ising  & -1 & 0 & exact solution \\
                            & $O(n)$ & -1 & 0  &  exact solution \\ 
 \hline
\multirow{2}{*}{$d=2$} & Ising  & -1.414$\ldots$ & 2.269$\ldots$ & exact solution\footnote{Exact values are $\varepsilon_c = -\sqrt{2}$ and $T_c = \frac{2}{\log(1 + \sqrt{2})}$.} \\
                            & $O(2)$ &  -1.4457(4) & 0.8929(1)  &  numerical \cite{Gupta,Hasenbusch2d} \\ 
 \hline
\multirow{3}{*}{$d=3$} & Ising  & -0.991(1) & 4.5112(3) & numerical \cite{3dIsing} \\
                            & $O(2)$ & -0.991884(6) & 2.2016(7)  &  numerical  \cite{3dXY} \\ 
                            & $O(3)$ & -0.9896(1) & 1.44298(2)  &  numerical \cite{3dHeis} \\ 
\end{tabular}
\end{ruledtabular}
\label{table_energies}
\end{table}

We now discuss known results, both analytical and numerical, in order to assess the validity of this conjecture in some particular cases. The results we were able to collect are reported in Table \ref{table_energies}. 
The conjecture is true for for systems with long-range interactions on $d$-dimensional lattices, $J_{ij} = N^{(\alpha/d) - 1}|i - j|^{-\alpha}$ with $0 \leq \alpha < d$; $\alpha = 0$ is the mean-field case of models defined on complete graphs with the same interaction strength between any two sites, $J_{ij} = 1/N$. All these systems have a mean-field-like phase transition at the maximum value of $\varepsilon$ ($\varepsilon_c = 0$ with our choice of units), with critical temperatures $T_c = 1/n$ \cite{Campa}. We stress again that critical energy densities are equal but critical temperatures are not and depend on $n$ \cite{quantum}. 
As to systems with nearest-neighbor interactions, the energy density range is $\varepsilon\in[-d,d]$ with our choice of units. The conjecture is true for $d=1$ at any $n$, although this case is somehow trivial because there is no transition at finite temperature. For $d = 2$, the Mermin-Wagner theorem rules out a long-range-ordered phase for any $n > 1$. However, a remarkable transition between a disordered and a quasi-ordered phase occurs for $n=2$ ($XY$ model), usually referred to as the Bere\v{z}inskij-Kosterlitz-Thouless (BKT) transition \cite{BKT}. 
In Table \ref{table_energies} we report the best recent estimate of the critical temperature obtained by Hasenbusch and coworkers (see e.g.\  \cite{Hasenbusch2d} and references quoted therein) and the corresponding critical energy density (estimated from a MonteCarlo simulation of a system with $256\times256$ spins \cite{Gupta}). The difference between this value and the exact value of the critical energy density of the Ising model on a square lattice is around $2\%$. This difference, though small, appears significant since it is orders of magnitude larger than the statistical error on the numerical estimate of the energy. Based on this result one should conclude that the conjecture is not verified in the case of the $XY$ model in $d=2$. However, we are comparing an exact result in the thermodynamic limit with a numerical estimate of the energy on a finite lattice, whose statistical accuracy does not consider the systematic error due to the finite size effects, which could be quite large in this particular case \cite{Leoncini,KennaIrving}. Moreover, also the precise determination of the critical temperature of the BKT transition is a subtle and difficult task due to its elusive nature.
This is witnessed by the remarkable spread of values of $T_c$ reported in different papers: the summary given in Ref.\ \cite{KennaIrving} shows that estimated critical temperatures vary in the interval $[0.88,0.99]$ while Ref.\ \cite{TobochnikChester} gave $[0.85,0.95]$ as confidence interval for $T_c$. The energy values given in Ref.\ \cite{Gupta} corresponding to both these temperature intervals do contain the Ising value $\varepsilon_c = -\sqrt{2}$; for instance, the temperature interval $[0.85,0.95]$ corresponds to $\varepsilon_c \in [-1.48,-1.38]$. We thus believe that the available data are not conclusive as far as a confirmation of the conjecture is concerned in this particular case. 

For $d=3$ the comparison is entirely between simulation outcomes, since no exact solution exists even for the Ising case. Results in Table \ref{table_energies} show that the critical energy measured for a $O(2)$ spin system ($XY$ model) \cite{3dXY} is clearly consistent with that measured for the Ising case \cite{3dIsing}. The difference between the estimated $\varepsilon_c$ of the $O(3)$ case (Heisenberg  model) \cite{3dHeis} and that of the Ising model is less than 1.5 times the error on the latter. Therefore the two estimates are consistent if one considers quoted errors as standard statistical errors. 

In conclusion, by considering a special class of stationary points of the energy landscape of $O(n)$ spin models we have proposed an approximate form of the microcanonical density of states and conjectured that the critical energy densities of $O(n)$ models with $n \geq 2$ equal those of the corresponding Ising models. Available analytical and numerical data are consistent with the conjecture, with the exception of the $O(2)$ case in $d=2$. The latter, however, seems the most interesting one. On the one hand, we already noted that the best recent estimates suggest the conjecture does not hold in this case but available data seem not conclusive. On the other hand, in this case the conjecture yields an exact prediction for the critical energy of the BKT transition in a model that is not exactly solvable. Therefore, in our opinion, more precise numerical estimates of the critical energy of the BKT transition would be very interesting, as well as estimates of other critical parameters made using the conjectured value of $\varepsilon_c$. Were the validity of the conjecture ruled out in the $XY$ case in $d=2$, it would confirm once more the special nature of the BKT transition and our conjecture might still be valid in a weaker form, i.e., restricted to $O(n)$ spin models with symmetry-breaking ferromagnetic phase transitions. Conversely, besides yielding an exact value for the critical energy of the BKT transition, a confirmation of the conjecture for the $XY$ model in two dimensions might hint at a stronger version of the conjecture. Such a stronger conjecture would be that any $O(n)$ spin model on a $d$-dimensional hypercubic lattice has a phase transition precisely at the critical energy density of the $d$-dimensional Ising model, even for $n > 2$ in $d=2$. We note that the presence of a phase transition in $O(n)$ models at any $n$ in two dimensions has already been suggested by Patrascioiu et al.\  \cite{Patrascioiu}. Although, to the best of our knowledge, no direct evidence of such a transition has been found yet, the possibility remains that the transition exists but is weak and elusive: our conjecture might help in finding it, suggesting where to look at.
Finally we note that our conjecture might be true for a completely different reason than the one argued in the present paper: the phase transitions in $O(n)$ models may all come from one of the finite-$N$ singularities of $\omega^{(n)}$ which survives as $N\to\infty$ because it becomes asymptotically flat, i.e., the determinant of the Hessian of ${\cal H}^{(n)}$ goes to zero. This is the condition for a finite-$N$ singularity to survive given in \cite{KSS}. However, preliminary results for $n=2$ and $d = 2,3$ \cite{tesi_Rachele} do not support this scenario.

\end{document}